\documentclass[runningheads]{llncs}
\usepackage[T1]{fontenc}
\usepackage{nameref}
\newcommand{\equalcontrib}{\textsuperscript{*}}
\newcommand{\EEGtpNine}{\texttt{RAW\_TP9}}
\newcommand{\EEGafSeven}{\texttt{RAW\_AF7}}
\newcommand{\EEGafEight}{\texttt{RAW\_AF8}}
\newcommand{\EEGtpTen}{\texttt{RAW\_TP10}}

\usepackage{graphicx}
\usepackage[style=numeric]{biblatex}
\usepackage{amsmath}
\usepackage{amssymb}
\usepackage[noabbrev,capitalize]{cleveref}

\usepackage{pdfpages}
\usepackage[normalem]{ulem} 
\usepackage{xcolor}

\begin{document}
%
\title{MEEDAV: A Synchronous Web Viewer for EEG, Eye-Tracking and Speech Data}
\titlerunning{MEEDAV: EEG, ET and Speech Viewer}
%
\author{Jan Pijálek\equalcontrib \and Karel Vlk\equalcontrib \and Ondřej Bojar\inst{1}\orcidID{0000-0002-0606-0050}
}
%

\authorrunning{J. Pijálek et al.}

\institute{ÚFAL, Faculty of Mathematics and Physics, Charles University, Czech Republic\\
\email{\{pijalek,vlk,bojar\}@ufal.mff.cuni.cz}\\
\url{https://ufal.mff.cuni.cz}}

\begingroup
\renewcommand\thefootnote{*}
\footnotetext{These authors contributed equally to this work.}
\endgroup

\maketitle              
\begin{abstract}
MEEDAV is an open-source web-based application for the synchronised visualisation of electroencephalography (EEG), eye-tracking, and audio data collected in psycholinguistic research. While originally developed for the Eyetracked Multi-Modal Translation (EMMT) corpus, which uses four-channel EEG data from the Muse 2 headband, MEEDAV also supports higher-density EEG setups thanks to its channel-agnostic processing pipeline. The system performs time alignment across all modalities and provides optional ICA-based EEG denoising. It features interactive Plotly visualisations, including unified EEG–audio–gaze timelines, gaze-intensity plots, event markers, and spatial heatmaps of fixation/saccade patterns. Researchers can filter by participant and stimulus, inspect raw versus cleaned signals, and compute cross-modal correlations. All processing is handled in real time, with a modular backend that supports local file access or GitHub-based streaming. Although initially tailored to the structure of the EMMT dataset, MEEDAV demonstrates a generalisable approach to multimodal data exploration and offers a lightweight, browser-accessible solution for cognitive neuroscience and translation studies.



\end{abstract}
%
%
%

\section{Introduction}
Understanding the cognitive processes underlying human language processing such as reading, speaking, translation, or generally language comprehension often requires the analysis of multimodal data, such as electroencephalography (EEG), eye-tracking, and audio. These data streams offer complementary insights into temporal dynamics, attention, and neural activity. However, aligning and visualising such heterogeneous signals in a coherent and interpretable manner presents both technical and methodological challenges.

To address this, we developed a web browser-based application for visualising synchronised EEG, eye-tracking, and audio recordings. Originally designed for internal analysis of the Eyetracked Multi-Modal Translation (EMMT) dataset \cite{ufal_emmt}, the tool enables researchers to explore temporal interactions between gaze behaviour and brain activity. The interactive interface allows filtering the dataset by participant, sentence, and stimulus type, and supports side-by-side inspection of raw and preprocessed signals.

Although the current implementation is tailored to the specific structure and format of the EMMT dataset, the application can be relatively easily adapted to other datasets and it provides several key features which could be inspiring for other researchers:

\begin{itemize}
    \item A compact simultaneous visualisation of EEG waveforms, a ``summary'' of eye movements using horizontal and vertical deltas, fixation/saccade events, and audio waveforms,
    \item Interactive gaze heatmaps for spatial visual attention,
    \item EEG denoising and spike removal via Independent Component Analysis (ICA, \cite{COMON1994287})
    \item Multi-panel views for comparing raw and cleaned signals.
\end{itemize}

All visual components are synchronised in time, facilitating visual identification of cross-modal correlations.

The application is implemented in Python, using Streamlit,\footnote{\url{https://github.com/streamlit/streamlit}} Plotly,\footnote{\url{https://github.com/plotly/plotly.py}} and Seaborn \cite{Hunter:2007, Waskom2021} for frontend rendering, and Pandas \cite{mckinney-proc-scipy-2010, reback2020pandas} for data manipulation. While the applicability is currently limited by the strict input format, the project serves as a proof-of-concept for accessible multimodal data exploration using open-source technologies.

This paper does not present a fully generalised solution, but rather documents the initial design and development process of the tool as an exploratory project output. Below, we briefly introduce the EMMT dataset.

\subsection{EMMT Dataset Overview}
The EMMT dataset comprises monocular eye movement recordings, audio data, and four-channel EEG signals collected from 43 participants. In each trial, participants were shown an English sentence, which they read aloud. Subsequently, they were presented with an image that sometimes provided contextual information. Participants were then asked to produce a Czech translation of the original sentence, possibly revising their initial version based on the content of the image.

Objective cognitive data were gathered using the Muse 2 EEG headband and the EyeLink 1000 Plus eye tracker.

The dataset is structured around two key units: \emph{stimuli}, each consisting of a sentence and a corresponding image, and \emph{probes}, which are sets of 32 stimuli grouped for experimental presentation. In total, the dataset contains 20 probes, provided in both raw form (as streamed from the acquisition hardware) and preprocessed form (time-aligned and cleaned).

\section{Related Work}
Several toolkits have been developed to facilitate the visualization of EEG data, often with support for additional modalities such as eye-tracking. However, for our purposes, many of these tools were either proprietary, lacked support for specific data types we needed to visualize (e.g., audio waveforms or eye-tracking events), or were too complex for our focused use case. Furthermore, since our team works across different platforms, deploying the application as a shared, web-based interface improved both usability and collaboration.

One related approach is EYE-EEG, an open-source MATLAB toolkit \cite{dimigen2011coregistration} designed to synchronize eye-tracking and EEG data using start and stop events. It also supports the detection of saccades and fixations. In our case, however, the EMMT dataset already provides synchronized data, with fixation and saccade events annotated by the eye tracker itself. Moreover, EYE-EEG does not natively support the visualization of additional data types such as audio or gaze displacement along the vertical and horizontal axes. Both EYE-EEG and our application use ICA to correct for ocular and other artifacts. However, our implementation performs this preprocessing automatically, while EYE-EEG allows for interactive artifact inspection and manual correction, which is substantially more tedious but should lead to higher precision.

Another comparable tool is OpenSync \cite{razavi2022opensync}, which focuses on real-time synchronisation of raw data streams during acquisition, primarily in neuroscience experiments. Since the EMMT dataset is already complete, our goals differ: we focus on post-hoc data exploration and preprocessing; areas that OpenSync does not address.
\section{System Design}
\label{sec:system}
\begin{figure}[t]
  \centering
  \includegraphics[width=\linewidth]{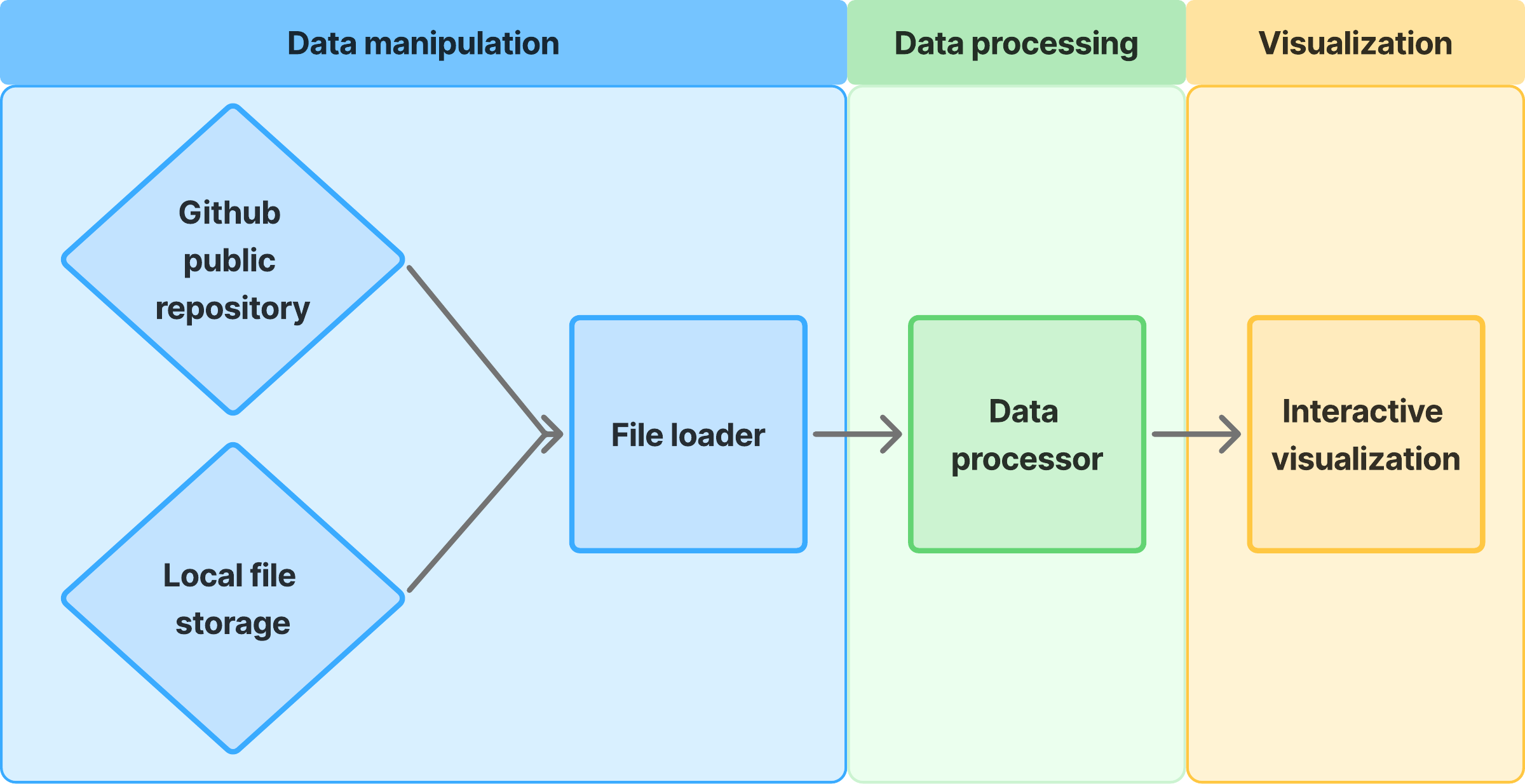}
  \caption{High-level data flow in MEEDAV. Blue = data manipulation,  
           green = processing, yellow = browser visualisation.}
  \label{fig:pipeline}
\end{figure}

This section details the architecture of our prototype, \textbf{MEEDAV}.\footnote{\textbf{M}ultimodal \textbf{E}EG, \textbf{E}ye-tracking, \textbf{A}udio \textbf{D}ata \textbf{A}nalyser \& \textbf{V}isualiser.} The application is organised into three semi-independent layers (\cref{fig:pipeline}).  
\cref{sec:overview} provides a high-level overview of the application, and its subsections explore the individual layers in greater detail. \cref{sec:features} then formally defines the principal interactive functionalities.

\subsection{Overview}
\label{sec:overview}

MEEDAV accepts EEG, gaze, and audio data files and processes them into a single time-aligned record.
The loader can resolve these artefacts either from a local folder or directly from the public repository over the GitHub REST API. Both backends implement the same interface so that switching storage is a one-line environment variable change.

\subsubsection{Data manipulation}
\label{sec:manip}

\paragraph{Why we adopted the EMMT layout.}
MEEDAV was conceived for exploratory work on the EMMT corpus, and therefore using the layout of the dataset's preprocessed folder was the most straightforward option.

\paragraph{Required modalities.}
Only the EEG stream is strictly required; eye-tracking and/or audio may be omitted.  
When present, the three files that together constitute a trial recording of the participant over a single stimulus must then have an identical basename (participant, sentence, order) and differ only in the modality suffix (e.g. \texttt{P03\_S084\_01\_Read.\{eeg, et, wav\}}).

Since the basename (\texttt{P03\_S084\_01\_Read}) functions as the primary key, it ensures that the EEG and gaze files are systematically associated with one another. When available, the corresponding audio file is also incorporated, resulting in a unified and coherent data unit. 

The original EMMT does not contain any preprocessed audio; only raw microphone recordings are provided.  
We therefore added a simple script that segments the raw stream per trial, saves each chunk as a 16-kHz mono WAV bearing the correct basename, and uploads the modified files to the same \texttt{preprocessed\discretionary{-}{-}{-}data\discretionary{/}{/}{/}\dots{}\discretionary{/}{/}{/}audio/} directory.  
Then, the application can properly use the newly created WAV files and optionally transform them further.

\paragraph{Local vs.\ GitHub backends.}
There are two subclasses, both deriving their functionality from the abstract \texttt{Loader} class:
\begin{itemize}
  \item \texttt{FileLoader}: opens paths from a local directory.
  \item \texttt{GithubFileLoader}: streams raw bytes via the REST endpoint\\
        \verb|/git/trees/main?recursive=1| and caches them in memory.
\end{itemize}

\paragraph{File formats.}
Both EEG and gaze logs follow CSV format. EEG headers follow Muse 2 channel names: \EEGtpNine{}, \EEGafSeven{}, \EEGafEight{}, \EEGtpTen{}.

EyeLink export fields are used in the gaze files: \verb|TimeStamp|, \verb|X|, \verb|Y|, \verb|Event|.

Audio is stored in the preprocessed WAV files. Since MEEDAV only needs the amplitude envelope for plotting and correlations, the reader returns a \texttt{BytesIO} stream and defers decoding to Librosa \cite{mcfee_2025_15006942}.

\subsubsection{Data processing}
\label{sec:proc}

Upon selecting a specific trial recording identified by a shared basename across EEG, eye-tracking, and audio files, the loader hands the data to a stateless \texttt{Data\-Processor} that:

\begin{enumerate}
  \item Converts wall-clock \texttt{TimeStamp} values into relative seconds.
  \item Up- or down-samples all modalities to a shared 256 Hz grid using linear interpolation.
  \item Normalises audio to the range $[-1, 1]$ and maps gaze coordinates to screen pixels.
\end{enumerate}

No heavy filtering is done here. Band-pass, ICA or artefact rejection are triggered lazily by the frontend and described later in \cref{sec:features}.  
The processor returns synchronised NumPy arrays and channel metadata; subsequent UI moves incur no further I/O.

\subsubsection{Data visualisation}
\label{sec:viz}

\begin{figure}[t]
  \centering
  \includegraphics[width=\linewidth]{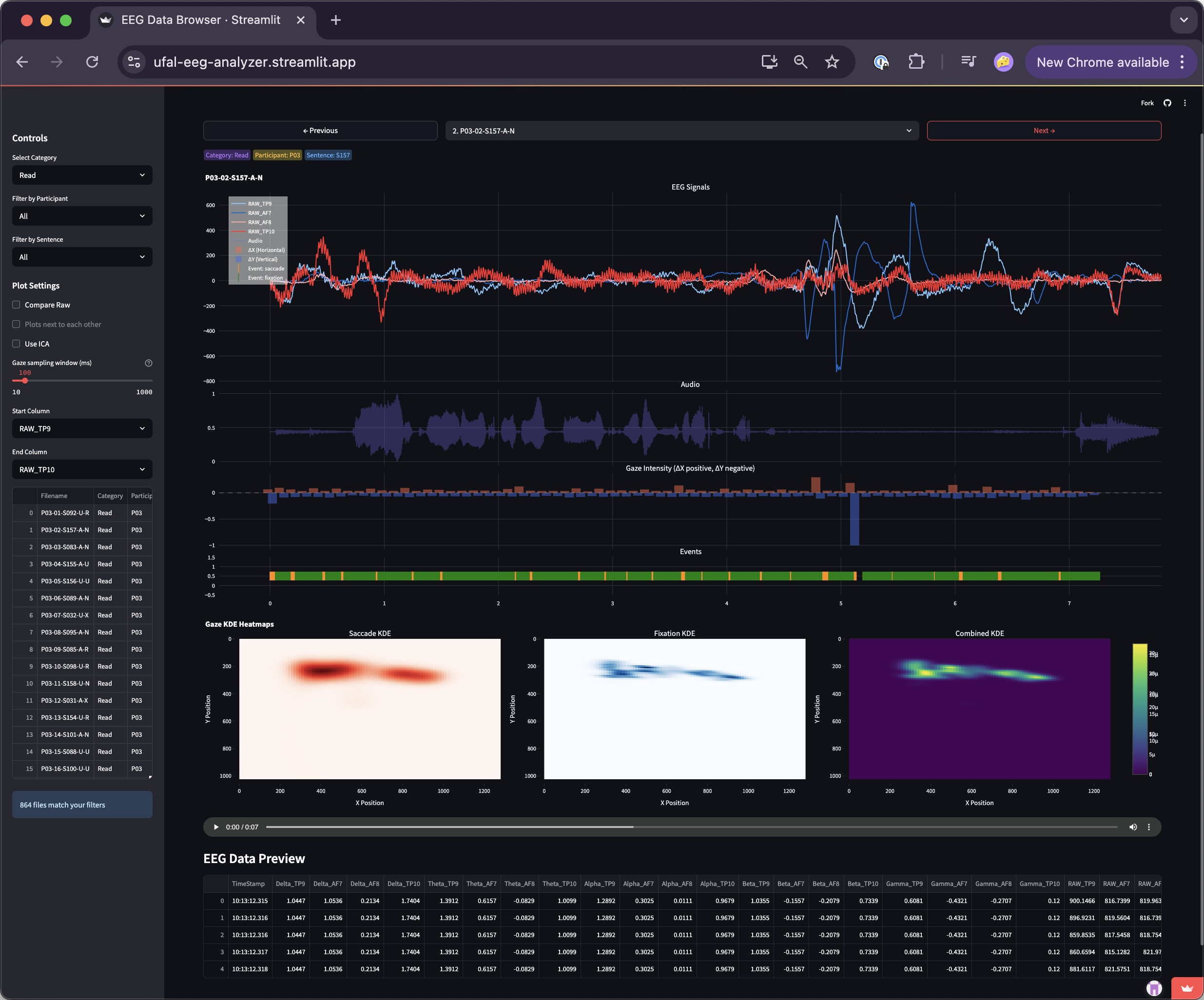}
  \caption{MEEDAV interface: synchronised EEG, audio, gaze-intensity bars, and
           event markers with unified hover cursor.  Tabs on the left toggle
           ICA-clean traces, KDE heat-maps, and the participant dashboard.
}
  \label{fig:meedav-ui}
\end{figure}

Our Streamlit-based frontend presents a four-row Plotly visualization comprising: EEG signals (four channels from the EMMT dataset), audio waveform, gaze-intensity bar plot, and eye-tracking event markers. The interface supports synchronized cross-modal navigation across all visualizations. Additional tabs provide kernel density estimate (KDE) heatmaps distinguishing fixations from saccades, and a participant dashboard (see Figure \ref{fig:meedav-ui}). While the default configuration supports four-channel Muse EEG, the plotting logic dynamically adapts to any channel list supplied by the processing module, ensuring compatibility with higher-density EEG data.

\subsection{Features and formal background}
\label{sec:features}

\paragraph{F1 — Gaze-intensity time-series.}
In order to represent the 2D motion of gaze in time in the two-dimensional plot where the $x$-axis represents time, we had to reduce the 2D gaze to one dimension. The EMMT stimuli have a particular form where the majority of horizontal eye movements corresponds to reading and the majority of vertical movements corresponds to comparison of text and an image displayed below it. We thus defined a conversion which highlights this semantics.

For every gaze sample $(x_t,y_t)$ we compute absolute motion magnitude as


\begin{equation}
m_t=\lvert x_t-x_{t-1}\rvert+\lvert y_t-y_{t-1}\rvert ,
\end{equation}

aggregate it over fixed windows $\mathcal B_k$, and normalize:

\begin{equation}
\tilde I_k=\frac{\sum_{t\in\mathcal{B}_k} m_t}{\max_j\sum_{t\in\mathcal{B}_j} m_t}.
\end{equation}

The implementation employs vectorized \texttt{pandas.diff()} with group-by summation and scaling. The resulting $\tilde{I}_k$ values are displayed as signed bar plots, where positive and negative bars encode horizontal and vertical components, respectively. Diagonal saccades yield symmetric peaks.

\paragraph{F2 — ICA-based EEG denoising.}
We expose the feature to suppress ocular and muscular artefacts in EEG signals using ICA. Given raw EEG data $\mathbf{X} \in \mathbb{R}^{C \times T}$, FastICA decomposes the signal under the model $\mathbf{X} = \mathbf{A} \mathbf{S}$, where $\mathbf{S}$ contains statistically independent sources and $\mathbf{A}$ is an unknown mixing matrix. Whitening is followed by iterative updates of the unmixing matrix $\mathbf{W}$ until $\mathbf{W}_{k+1}^\top \mathbf{W}_k = \mathbf{I}$,
yielding source estimates $\hat{\mathbf{S}} = \mathbf{W} \mathbf{X}$. Artefactual components are identified by high kurtosis ($|\mathrm{kurtosis}| > 3\sigma$) or large peak-to-peak amplitude (above the 95th percentile) and are set to zero. The denoised signal is reconstructed as $\tilde{\mathbf{X}} = \mathbf{A} \hat{\mathbf{S}}$. The process completes in under 150\,ms for a 30\,s segment with 4 channels.

\paragraph{F3 — Multi-modal timeline.}
A unified Plotly figure is created within the function
\texttt{processor\discretionary{.}{.}{.}create\discretionary{\_}{\_}{\_}interactive\discretionary{\_}{\_}{\_}plot()}.
EEG traces are plotted with a colour map that greys out invalid channels based on peak-to-peak amplitude heuristics.
Audio signals are amplitude-normalised ad-hoc and plotted as continuous waveforms along the timeline.  
Event markers are vertical lines coded by type (saccade, fixation).

\paragraph{F4 — Spatial KDE heat-maps.}
Given fixation $(x_i, y_i)$ or saccade landing points, a two-dimensional Gaussian kernel density estimate (KDE) $f(x, y)$ is computed over a $100 \times 100$ grid.  
This grid spans the full screen resolution used in the EMMT dataset experiments, with horizontal and vertical dimensions set to $x_{\text{max}} = 1280$ and $y_{\text{max}} = 1024$, respectively.  
The resolution of 100 sampling points per axis is chosen to balance spatial detail with computational efficiency.  
The resulting density maps are visualized using one of three colour maps: \texttt{Reds}, \texttt{Blues}, or \texttt{Viridis}.

\paragraph{F5 — EEG audio/gaze correlations.}
Signals are resampled to 10 ms resolution, and Pearson, Kendall, or Spearman correlations are computed within sliding time windows.  
For each window, the correlation coefficient is given by:

\begin{equation}
r_{xy} = \frac{1}{N - 1} \sum_i \frac{(x_i - \bar{x})(y_i - \bar{y})}{\sigma_x \sigma_y}.
\end{equation}

Per-channel results are returned as a Python dictionary and displayed as sortable bar plots in the UI.
\section{Conclusion}
\subsection{Final Remarks}
We presented MEEDAV, a web-based tool for the multimodal exploration of synchronised EEG, eye-tracking, and audio data. Designed for use with the EMMT dataset, MEEDAV provides an accessible, time-aligned interface combining signal plots, gaze heatmaps, and ICA-denoised EEG visualisation. The system architecture is modular and adaptable, using open-source Python libraries and supporting both local and GitHub-based data access.

\subsection{Future Work}
As noted earlier, the current implementation closely follows the specific file structure and naming conventions of the original EMMT dataset. To make the tool more flexible and reusable, we plan to introduce configurable file path and name pattern mappings, allowing users to adapt the application to other datasets with minimal overhead.

Another limitation lies in the restricted input format support: MEEDAV currently handles only CSV-like data structures for EEG and eye-tracking signals. This forces users to manually convert their data from other common formats. In future versions, we aim to support additional EEG file formats such as FIF or Artemis123, leveraging existing libraries like MNE to enable broader compatibility.

Finally, the frontend performance remains a significant bottleneck, primarily due to the computational and memory demands of Plotly-based visualizations. We are currently exploring a migration to Altair-Vega Lite \cite{VanderPlas2018}, which promises improved rendering efficiency and better scalability across devices.




%
%
%
%
\printbibliography[heading=bibintoc]
\end{document}